# Stacking tunable interlayer magnetism in bilayer CrI$_3$

Peiheng Jiang[1,†], Cong Wang[2,†], Dachuan Chen[1], Zhicheng Zhong[1,*], Zhe Yuan[3], Zhong-Yi Lu[2] and Wei Ji[2,*]

[1] *Key Laboratory of Magnetic Materials and Devices & Zhejiang Province Key Laboratory of Magnetic Materials and Application Technology, Ningbo Institute of Materials Technology and Engineering, Chinese Academy of Sciences, Ningbo 315201, P.R. China*

[2] *Beijing Key Laboratory of Optoelectronic Functional Materials & Micro-Nano Devices, Department of Physics, Renmin University of China, Beijing 100872, P.R. China*

[3] *The Center for Advanced Quantum Studies and Department of Physics, Beijing Normal University, Beijing 100875, P.R. China*

*Corresponding authors: W.J. (email: wji@ruc.edu.cn) and Z.Z. (email: zhong@nimte.ac.cn)*

*† These authors contributed equally to this work.*

Diverse interlayer tunability of physical properties of two-dimensional layers mostly lies in the covalent-like quasi-bonding that is significant in electronic structures but rather weak for energetics. Such characteristics result in various stacking orders that are energetically comparable but may significantly differ in terms of electronic structures, e.g. magnetism. Inspired by several recent experiments showing interlayer anti-ferromagnetically coupled CrI$_3$ bilayers, we carried out first-principles calculations for CrI$_3$ bilayers. We found that the anti-ferromagnetic coupling results from a new stacking order with the C2/m space group symmetry, rather than the graphene-like one with $R\bar{3}$ as previously believed. Moreover, we demonstrated that the intra- and inter-layer couplings in CrI$_3$ bilayer are governed by two different mechanisms, namely ferromagnetic super-exchange and direct-exchange interactions, which are largely decoupled because of their significant difference in strength at the strong- and weak-interaction limits. This allows the much weaker interlayer magnetic coupling to be more feasibly tuned by stacking orders solely. Given the fact that interlayer magnetic properties can be altered by changing crystal structure with different stacking orders, our work opens a new paradigm for tuning interlayer magnetic properties with the freedom of stacking order in two dimensional layered materials.

*Introduction.-* Magnetism in two dimensions has received growing attention since the two ferromagnetic monolayers, namely $CrI_3$ [1] and $Cr_2Ge_2Te_6$ [2], were successfully fabricated in 2017. The ferromagnetism in these two layers was believed to be stabilized by magnetic anisotropy as enhanced by spin-orbit coupling or external magnetic fields. Their Curie temperatures were up to ~50 K. Very recently, a room-temperature $T_c$ were achieved in monolayer $VSe_2$ [3] and $MnSe_x$ [4], two members of the transition-metal dichalcogenides family. This shed considerable light on the search for high $T_c$ ferromagnetic (FM) magnets. However, the tunability of magnetism has been emerging as a new challenge. The coupling strengths of two-dimensional (2D) materials are significantly different between intra- and inter-layer interactions. Such difference may offer diverse magnetic coupling mechanisms at strong and weak interacting limits. The interlayer magnetic coupling is of peculiar interest, as the effective coupling is relatively weak and confined within few atomic layers, which is much easier to model and more feasible to tune than strong and periodic couplings in three-dimension.

Recent experiments demonstrated that the anti-ferromagnetic (AFM) interlayer order in bilayer $CrI_3$ can be manipulated to a FM order by electric gating or reasonably large magnetic fields [5-12]. As a consequence, a magnetic tunnel junction with giant tunneling magnetoresistance values was achieved in bilayer $CrI_3$ devices [5-8]. These experimental demonstrations may open a new avenue for "interlayer" spintronics in magnetic bilayers. There are a few conjectures for the mechanism of the magnetic tunability, however, these arguments lack compelling supports that even the details of the stacking geometry and the magnetic ground state are yet to be addressed [11]. The tunability also strongly relies on the initial geometry and the associated magnetic ground state of the bilayers, from which external fields change the magnetism. The interlayer stacking order was manifested as an effective and sustained way for tailoring geometry and the accompanying properties of bilayers, e.g. five times reduced shear force constants [13] and emerged strong correlation of electrons [14] in twisted graphene, unusual optical signals in folded $MoS_2$ [15] and a band tail state observed in simple-sliding $MoSe_2$ bilayers [16]. In light of this, it seems paramount to unveil the ground state stacking order and its roles in varying interlayer magnetic couplings and in selecting magnetic ground state of $CrI_3$ bilayers.

Here, we carried out first-principles calculations to unveil the stacking-dependent interlayer electronic and magnetic couplings in the $CrI_3$ bilayer. The intralayer FM of $CrI_3$ was ascribed to a Cr-I-Cr FM super-exchange in which the Cr-I-Cr bond angle approaches 90° [17]. As a result of the Hund correlation effect, the magnetic moments of the both Cr atoms align parallel, which is rather robust under external perturbations. In terms of interlayer magnetism, a simple sliding of one layer of the bilayer could change the direct hopping strength between interlayer I orbitals, which varies the interlayer magnetic ground state of the bilayer between interlayer FM and AFM ones. The AFM ground state allows magnetic-field to control tunnel magnetoresistance effect, which was realized in the stacking tuned AFM coupled $CrI_3$ bilayer.

*Method.-* Our density functional theory calculations were performed using the generalized gradient approximation and the projector augmented wave method [18] as implemented in the Vienna *ab-initio* simulation package (VASP) [19]. Dispersion correction with the optB86b functional [20] was adopted for structure related calculations. The optimized lattice constants were explicated shown in supplementary Table SI. For energy comparisons among different magnetic configurations, we used the PBE or HSE06 functional, with the inclusion of spin-orbit coupling (SOC), based on the vdW-DF revealed structures. On-site Coulomb interaction to the Cr *d* orbitals was considered with a U value of 3.9 eV and a J value of 1.1 eV. The adoption of different functionals, modification of the interlayer distance, stacking of graphene or BN layers to the bilayer, and different U values have been checked and all calculations support our conclusions which were provided in the supplementary materials [21].

*Results and discussion*.- Bulk CrI$_3$ exhibits a van der Waals structure and possesses a rhombohedral structure with the $R\overline{3}$ space group symmetry at low temperature (the LT phase). When temperature increases to 210 K - 220 K, it undergoes a structural phase transition to a monoclinic lattice with the C2/m space group symmetry (the HT phase) [22]. It is expected that a bilayer CrI$_3$ has similar structures to its bulk counterpart, namely, rhombohedral and monoclinic structures for low- and high-temperature exfoliated CrI$_3$ bilayers, respectively. Figure 1(a-b) shows the structures of the LT and HT phases of the CrI$_3$ bilayer, respectively.

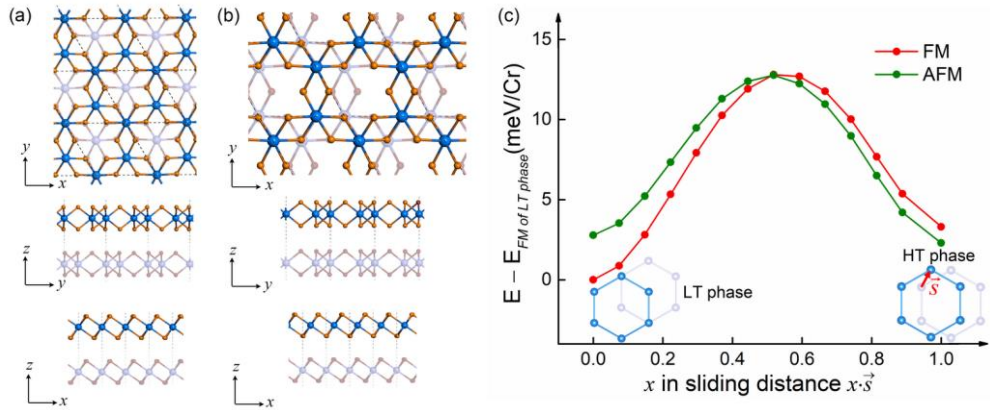

Figure 1 (a-b) Top and side views of the CrI$_3$ bilayer in the low temperature (LT) (a) and high temperature (HT) (b) phases. Slate-blue and dodger-blue balls represent Cr atoms and orange and maroon balls for I atoms. (c) Transition pathways between the two phases in FM and AFM configurations. The insets schematically illustrate LT and HT phases, indicating how interlayer structure changes during the transition between these two phases. Spin-orbit coupling was not included for plotting panel (c).

The structural difference between these two phases of CrI$_3$ bilayers can be viewed as different stacking orders of single CrI$_3$ layers. In the LT phase, the stacking order is in analogue to an AB-stacked graphene or a 2H-phase MoS$_2$ bilayer, in which a Cr atom of the bottom layer sits below the hollow site of the Cr hexagon as shown in Fig. 1(a). The LT phase is roughly 2.35 meV/Cr lower in energy than the HT phase. The metastable HT phase can be viewed as sliding the upper layer from the previous LT position with vector $\vec{s}$ as indicated in Fig. 1(c). Figure 1(c) also illustrates the transition pathways from the LT to the HT phase in either interlayer FM or AFM coupled configuration (see detail in Supplementary Figure S1). This indicates a transition barrier of roughly 10 meV/Cr, which may prevent the metastable HT phase transferring to the LT phase in the CrI$_3$ bilayer.

Magnetic ground state plays an important role in determining physical properties of materials. Experimentally, CrI$_3$ monolayer has a very strong intralayer FM order that persists up to ~50 K [1]. Consistently, in our calculation, the FM state is at least 12 meV/Cr more favored than other magnetic configurations. Given this highly stable intralayer FM ground state, we next considered the interlayer magnetic couplings of the CrI$_3$ bilayer by comparing its energies with interlayer FM and AFM configurations. It is exceptional that the interlayer FM state is 3.23 meV/Cr more stable than the interlayer AFM state in the LT phase. Such an energy difference is less influenced under different on-site Coulomb U values (see Supplementary Figure S2). This interlayer FM ground state is so robust that it is unlikely to be altered by general manipulation methods, such as strain, electric field, doping and among the others (see Supplementary Figure S3 and S4).

In terms of the HT phase, however, the interlayer magnetic ground state is an AFM one [7] with an energy difference ($E_{AFM}$-$E_{FM}$) of -0.54 meV/Cr. The change of U value, adoption of different functionals, modification of the interlayer distance, or stacking of graphene or BN layers to the bilayer varies the exact relative energy but does not change the order of stability of these two magnetic configurations (see Supplementary Figure S2, S4, S5, Table SII, and Table SIII). Such a small energy difference implies the interlayer spin-exchange coupling is rather weak ($J$~0.5 meV) in the HT phase, although the intralayer magnetic coupling was found much stronger ($J$~3 meV). The weak interlayer magnetic coupling is consistent with the facts that the bandgap of the bilayer varies less than 0.15 eV from that of a CrI$_3$ monolayer [23], and that the cohesive energy of the bilayer is relatively small with a value of 14 meV/Å$^2$. Such a weak magnetic coupling in the HT phase indicates that the manipulation of its interlayer magnetism is, most likely, feasible by applying an external magnetic field.

These results of the LT and HT phases suggest that the interlayer AFM coupled CrI$_3$ bilayer [5,6,8-12], could be, most likely, maintained in the HT phase, rather than the presumed LT phase, even at low temperatures. This is, we believe, ascribed to structural quenching under rapid cooling rates and/or vertical confinement from the capping layers in the measurements, which could be directly verified by control experiments with slow cooling rates and removed capping layers. We additionally examined the responses of the both phases to electric field and charge doping. The HT phase results are highly consistent with the measurements [11], i.e., unremarkable effects from hole doping and a transition from interlayer AFM to FM coupling resulted from electron doping, while those of the LT phase are against to the experiments (supplementary Fig. S3 and S6); this further verifies the solidness of our conclusions.

We further extended our calculations to CrI$_3$ tri- and quad-layers to investigate whether the interlayer AFM order also maintains in CrI$_3$ multilayers. The interlayer AFM state still holds in all the considered HT multilayers and its bulk counterpart as shown in Fig. 2, which fully coincide with the recent experimental results [5,6]. This consistency strongly supports that the CrI$_3$ multilayers measured in the experiments still maintain the HT phase even at low temperature. In terms of the LT multilayers and the bulk form, the FM state is always the interlayer magnetic ground state.

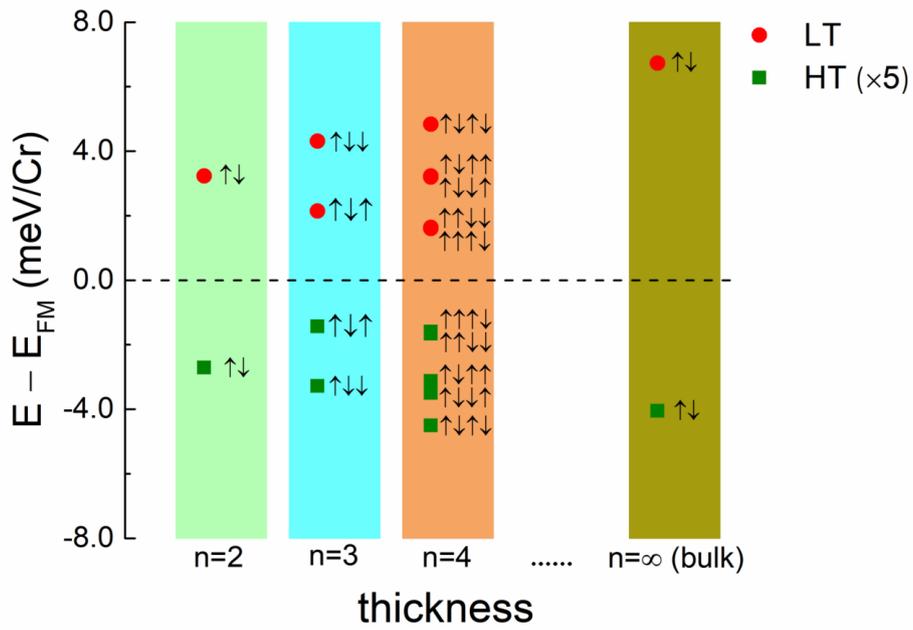

Figure 2 Energy of different interlayer magnetic orders for bilayer (n=2), tri-layer (n=3), quad-layer (n=4) and bulk (n=∞) $CrI_3$. The symbol ↑ and ↓ represent spin up and spin down, respectively. All energies have subtracted the energies of ferromagnetic (FM) states. The red and olive dots represent the energy of LT and HT phases, respectively. The energy of HT phase has been enlarged 5 times in order to show clearly. Here, the spin-orbit coupling has been considered.

Given the established magnetic ground state, we carefully examined the stacking difference resulted from variation of interlayer magnetic couplings in the $CrI_3$ bilayer. Figure 3(a) shows the charge accumulation at the interlayer region after stacking two $CrI_3$ layers together in the LT phase. Here, the amount of redistributed charge is comparable with that previously found in $MoS_2$ [24] (see Supplementary Figure S7). The accumulated charge mainly resides between close-contacted (4.20 Å) interlayer I-I pairs. The red arrow indicates a I-I pair that shares an appreciable amount of electrons, which bridges the interlayer magnetic coupling as we elucidated below.

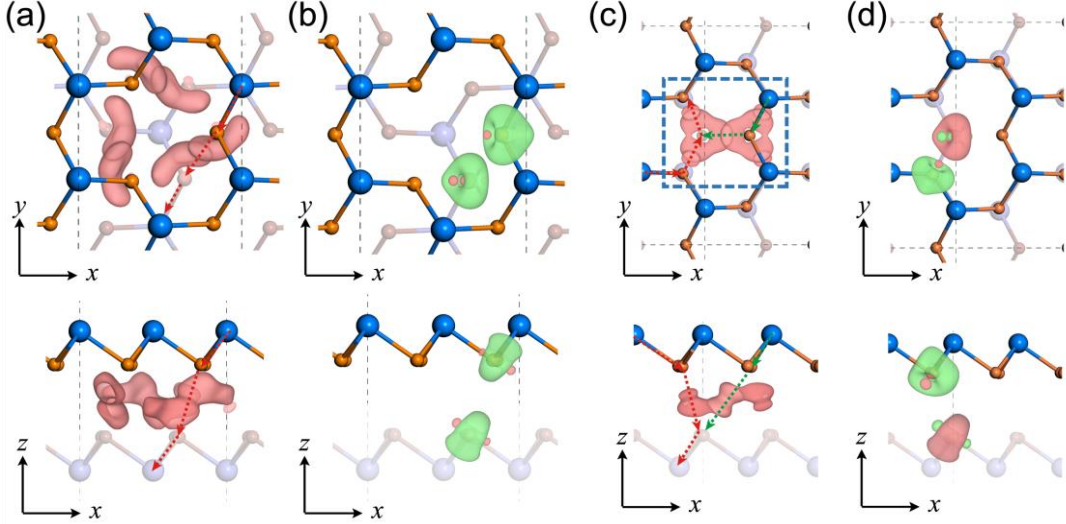

Figure 3 Mechanism of interlayer magnetic couplings in LT and HT phases. (a) Differential charge density of the $CrI_3$ bilayer in the LT phase with both intra- and interlayer FM order (isosurface value of 0.0001 $e$/Bohr$^3$). Here, the light-rose isosurface contours show the charge accumulation after stacking those two layers together. Red dashed arrows indicate two interacting I atoms from the two $CrI_3$ layers. (b) Spin density of the two mentioned I atoms marked in (a). Red and green isosurface contours correspond to spin-up and -down charge densities, respectively. (c) Differential charge density of the $CrI_3$ bilayer in the HT phase with intralayer FM and interlayer AFM orders. The red and green dashed arrows, again, show the two interacting I atoms from both layers. (d) Spin density of the I atoms marked in (c).

In particular, the intralayer FM coupling is through a FM Cr-I-Cr super-exchange, in which the Cr-I-Cr bond angle (93°) approaches 90° [17]. Here, we defined the two orbitals involved in the intralayer super-exchange as $p_x$ and $p_y$. As shown in supplementary Fig. S8. Both Cr (3.28 $\mu_B$) and I (-0.12 $\mu_B$) have local magnetic moments and the moments are in opposite directions, suggesting the spins of electrons of I atoms are polarized by Cr atoms. Local geometry shows that each I atom of the marked interlayer I-I pair has a $p_{x/y}$ orbital interacting with the other one. Both orbitals are in a nearly linear configuration (an angle of 160°), indicating that the interaction is not owing to a FM super-exchange. However, the shared electron indicates direct hybridization between the two $p_{x/y}$ orbitals, leading to a direct FM coupling between the spin-down components (green) of the two I atoms. This mechanism is supported by the charge reduction around the $p_{x/y}$ orbitals and the charge accumulation at the interlayer region in the spin-dependent differential charge density (DCD) of the spin-down component, as shown in Supplementary Fig. S8. The FM coupled interlayer I atoms thus give rise to FM coupled interlayer Cr atoms through the intralayer Cr-I magnetic hybridization.

Here, both intralayer and interlayer Cr atoms are FM coupled, but are with different mechanisms. The both couplings are at the two extremes in terms of interacting strength that the intralayer Cr-I-Cr super-exchange sits at the strong-interaction limit while the interlayer Cr-I…I-Cr direct exchange lies in the weak-interaction limit. Direct exchange couplings were usually found in metals, but in this case, it was found in a semiconductor with a non-metal element. This exception is, most likely, due to strongly extended $p$ wavefunctions of I and the vdW attraction induced overlap of interlayer $p_{x/y}$ orbitals. Although it is much weaker, the overlap is also a result of known covalent-likely quasi-bond, as revealed in other 2D materials [24-28]. A similar but much stronger interlayer magnetic coupling ($J$~10 meV) was found in $CrS_2$ bilayers [29], in which the bilayer strongly favors interlayer FM coupling and even changes an intralayer AFM order to the FM order. These results suggest that a much weaker interlayer coupling allows the interlayer magnetism to be more feasibly tuned.

Given the much weaker interlayer magnetic interaction of $CrI_3$, we thus expect the interlayer FM could be tuned to AFM through external perturbations. Given that a minor $p_z$ spin density shows opposite sign to that of $p_{x/y}$ (Fig. 3(b)), a straightforward idea is to shift one layer of the bilayer, which favors direct exchange of a $p_{x/y}$ orbital from one layer with a $p_z$ orbital from the other layer. The HT bilayer is just the case for this idea. Figure 3(c) shows the interlayer DCD of the HT bilayer. It shows the charge redistribution is slightly weaker than that of the LT bilayer and the charge accumulation mainly occurs around the six I atoms as marked in the blue dotted rectangular frame. In the HT case, electron sharing is not within a I-I pair, but through a tri-I cluster forming a triangle-shaped accumulated charge density. A red dashed arrow (Fig. 3(c)) marks two interlayer I atoms in the left triangle. The Cr-I…I-Cr interaction does not take a nearly linear configuration but is in a 135° configuration that a $p_{x/y}$ orbital of the top-layer I atom is oriented toward a $p_z$ orbital of the bottom layer I atom with an interlayer I-I distance of 4.20 Å (red arrows). Direct charge transfer from the $p_{x/y}$ orbital to the $p_z$ orbital is observable in the spin-dependent DCD, indicating a direct $p_{x/y}$-$p_z$ interaction. Here, the spin-orbit coupling may play a key role that the different parity symmetries of the $p_{x/y}$ and $p_z$ orbitals do not obscure the hybridization. Since the $p_{x/y}$ orbital is spin-polarized in the opposite direction to the $p_z$ orbital, the direct $p_{x/y}$-$p_z$ interaction results in an AFM coupling between interlayer I atoms. On the other hand, charge transfer between two triangles also gives rise to a $p_{x/y}$-$p_{x/y}$ interaction (green arrows) with a larger distance of 4.31 Å, making a weaker FM coupling compared with the former AFM one. Two competing Cr-I…I-Cr interactions, a stronger AFM $p_{x/y}$-$p_z$ interaction and a weaker FM $p_{x/y}$-$p_{x/y}$ interaction, coexist in HT bilayer $CrI_3$, leading to the AFM coupled bilayer with a smaller FM/AFM energy difference.

*Discussion and conclusions.-* In summary, we have carefully investigated the stacking orders of $CrI_3$ bilayers and successfully revealed stacking dependent magnetic couplings between the two $CrI_3$ layers. There are, at least, two stacking orders for $CrI_3$ bilayers, namely the HT and LT bilayers. Both bilayers take the FM intralayer magnetic order and the HT bilayer favors an AFM interlayer magnetic state. This AFM interlayer coupling is nearly decoupled from the intralayer FM coupling and the strength of it is largely reduced. In light of this, the interlayer magnetic configurations are changeable by a reasonably large external field, perturbations or change of local stacking geometry while the intralayer FM state still maintains. Moreover, we should indicate that various stacking orders of $CrI_3$ bilayers shall be experimental accessible at low temperature, given the three experimentally realized stacking orderings of $MoS_2$ using well controlled fabrication conditions [30]. This offers an effective way to tune interlayer magnetic configuration in $CrI_3$ bilayers because of a locked stacking-magnetism coupling.

In previously revealed 3D magnetic materials, large magnetic moments and strong spin-exchange coupling are usually paramount to resist thermal fluctuation thereby achieving a high Curie temperature for practical applications. However, such large moments with strong spin-exchange couplings result in a significant amount of energy needed to manipulate the magnetic moments. Here, in 2D $CrI_3$ layers, the strong intralayer FM coupling keeps the magnetic moments ordered within each layer at finite-temperature, yet the weak interlayer AFM coupling in the HT bilayer allows the magnetic moment of each layer being feasibly manipulated. Given such close energies of the two couplings, we infer that magnetic domains with either FM or AFM interlayer coupling may be observable in a large area bilayer $CrI_3$ sample, which calls for subsequent experiments to verify. In addition, these two nearly decoupled magnetic couplings governed by two different mechanisms at the two interaction limits combines two apparently conflicting requirements of magnetic materials, which points to a new direction for seeking magnetic 2D layers in real applications. In addition, after being shown in tailoring mechanical, optical and electrical properties, layer stacking was also illustrated of magnetic tunability.

*Acknowledgements.-* We thank Dr. Jiadong Zhou at Nanyang Technological University of Singapore for valuable discussions. We gratefully acknowledge financial support Project supported by the National Natural Science Foundation of China (Gant Nos., 11622437, 61674171, and 11774360), the Fundamental Research Funds for the Central Universities, China, and the Research Funds of Renmin University of China (Grant No. 16XNLQ01). C.W was supported by the Outstanding Innovative Talents Cultivation Funded Programs 2017 of Renmin University of China. Calculations were performed at the Physics Lab of High-Performance Computing of Renmin University of China, Shanghai Supercomputer Center and Supercomputing Center of Ningbo Institute of Materials Technology and Engineering.

# Supplementary Materials

# for

# "Stacking tunable interlayer magnetism in bilayer CrI$_3$"

Peiheng Jiang[1,†], Cong Wang[2,†], Dachuan Chen[1], Zhicheng Zhong[1,*], Zhe Yuan[3], Zhong-Yi Lu[2] and Wei Ji[2,*]

[1] *Key Laboratory of Magnetic Materials and Devices & Zhejiang Province Key Laboratory of Magnetic Materials and Application Technology, Ningbo Institute of Materials Technology and Engineering, Chinese Academy of Sciences, Ningbo 315201, P.R. China*

[2] *Beijing Key Laboratory of Optoelectronic Functional Materials & Micro-Nano Devices, Department of Physics, Renmin University of China, Beijing 100872, P.R. China*

[3] *The Center for Advanced Quantum Studies and Department of Physics, Beijing Normal University, Beijing 100875, P.R. China*

*\*Corresponding authors: W.J. (email: wji@ruc.edu.cn) and Z.Z. (email: zhong@nimte.ac.cn)*

*† These authors contributed equally to this work.*

# Calculation method:

In our density functional theory calculations, the uniform Monkhorst-Pack **k** mesh of 15×15×1 was adopted for integration over Brillouin zone. The lattice constants were optimized with both FM and AFM states, respectively (see supplementary Table SI). A plane-wave cutoff energy of 700 eV was used during the structural relaxations. A sufficiently larger distance of $c > 15$ Å along out-of-plane direction was adopted to eliminate interaction between each layer. Dispersion correction was made at the van der Waals density functional (vdW-DF) level, with the optB86b functional for the exchange potential, was adopted for structure related calculations. The optimized lattice constants were explicated shown in supplementary Table SI. Other several vdW functionals (see supplementary Table SII) were also checked and all of them give the same conclusion which verify the reliability of our results. A self-consistently calculation of the U value based on a linear response method gives U=3.9 eV and J=1.1 eV, which were used in our calculations. Spin-orbit coupling (SOC) was considered in all comparison of energies for different magnetic configuration, in which the PBE or HSE functional was used rather a the vdW-DF functional used in structural relaxations.

We should indicate that most of calculations of FM and AFM states were done with the same structure (the one optimized with FM state, as done in the main text). The reason should be that the energy difference between FM and AFM states is quite small, any difference in structure may lead to an energy difference comparable with it. In order to ensure that the energy difference comes from the magnetic configuration instead of structure, we adopt the same structure in both calculation of FM and AFM state.

The comparison of FM and AFM energies with different structures, i.e., AFM energies with AFM relaxed structures and FM energies with FM relaxed structures, have also been done which give similar results with that of same structural calculations (see supplementary Fig. S2).

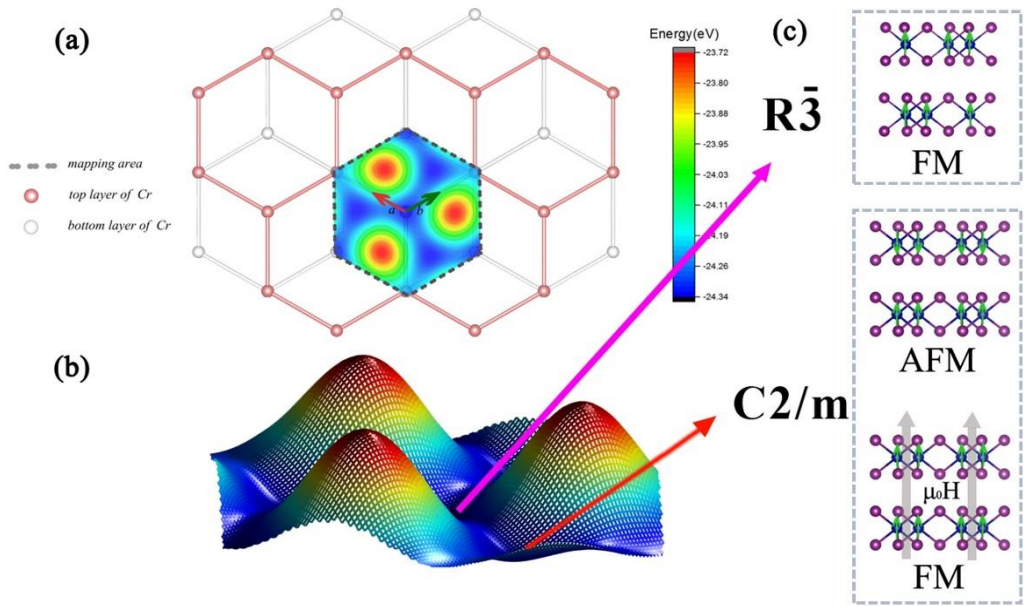

Figure S1 Scanning of stacking order in bilayer CrI$_3$. Different stacking orders can be obtained by fixing the bottom layer (represented by grey atoms) and moving the top layer (represented by light-rose atoms) as shown in Fig. S1(a). The total energies of bilayer CrI$_3$ with different stacking orders are denoted by color and height in Fig. S1(a) and 1(b), respectively. Two (meta)stable stacking orders, namely $R\bar{3}$ (the LT phase) and C2/m (the HT phase) phases, are found, which are indicated by blue area in Fig. S1(a) and energy valleys in Fig. S1(b). The *z*-axis of every atom in all structural optimizations have be fully relaxed. Fig. S1(c) shows the atomic structures of $R\bar{3}$ and C2/m phases. The ground states of $R\bar{3}$ and C2/m phases are FM and AFM, respectively. Meanwhile, the AFM state of C2/m phase can be tuned to FM state under a reasonable external magnetic field which coincides very well with recent experiments. It indicates that the AFM CrI$_3$ bilayer may be in the C2/m phase, rather than the $R\bar{3}$ phase as previously believed. On the other hand, the same magnetic configurations are also predicted in CrI$_3$ multilayer and bulk form with the HT and LT phases, respectively. Therefore the same intrinsic mechanism is expected and their magnetic configurations can be also tuned by the same measurements as that of the bilayer one.

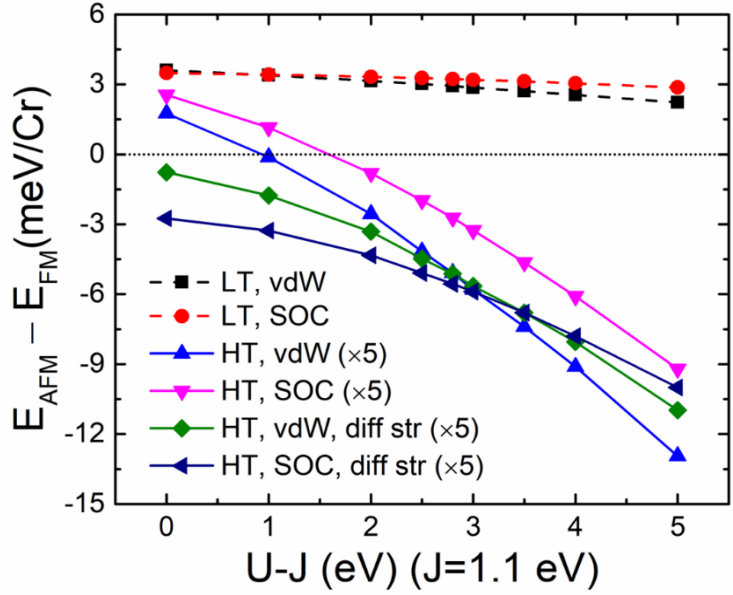

Figure S2 Energy differences between AFM and FM states ($E_{AFM}$-$E_{FM}$) under different on-site Coulomb U-J (J was fixed to 1.1 eV) values for the LT and HT phases. The energy differences of HT phase have been enlarged 5 times. For the LT phase (black and red dots for PBE+vdW and PBE+SOC approaches, respectively), the energy differences change less under different U-J values hence the interlayer FM ground state is always maintained. When it comes to the HT phase, interlayer AFM state becomes energetically more stable with U-J larger than 1.0 eV for the PBE+vdW approach (magenta dots) and 1.5 eV for the PBE+SOC approach (blue dots). Above calculations were done with the same structure for FM and AFM states (both with the structure optimized with FM state, see detail in supplementary calculation method). When different structures (denoted as "diff str") of FM and AFM are adopted (optimized with FM and AFM states respectively, see detail in supplementary calculation method), for both PBE+vdW (olive dots) and PBE+SOC (navy dots) approaches, the AFM states of HT phase are more stable than FM states no matter which value of U is used. In our calculations, the U=3.9 eV and J=1.1 eV (U-J=2.8 eV) was adopted which is calculated using linear response method and close to the value reported in the literatures (PRB 97, 245409 (2018)). Difference schemes of on-site Coulomb U (Phys. Rev. B 52, R5467 (1995); Phys. Rev. B 57, 1505 (1998)) have been checked that all of them give similar results.

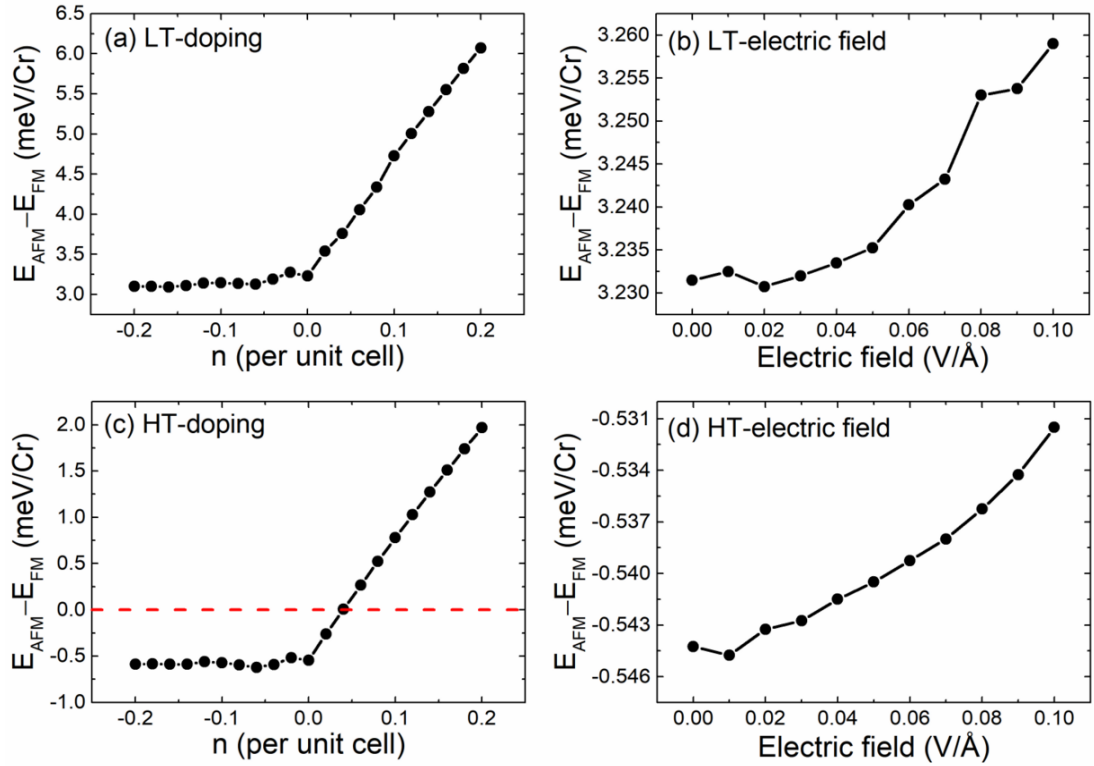

Figure S3 Energy differences between AFM and FM states ($E_{AFM}$-$E_{FM}$) for the LT phase and HT phase under doping and electric field. In the LT phase, the interlayer FM ground state is so robust that it is unlikely to be altered by doping and electric field. While the electron doping can easily tune the HT phase from interlayer AFM to FM configuration. Spin-orbit coupling was considered here.

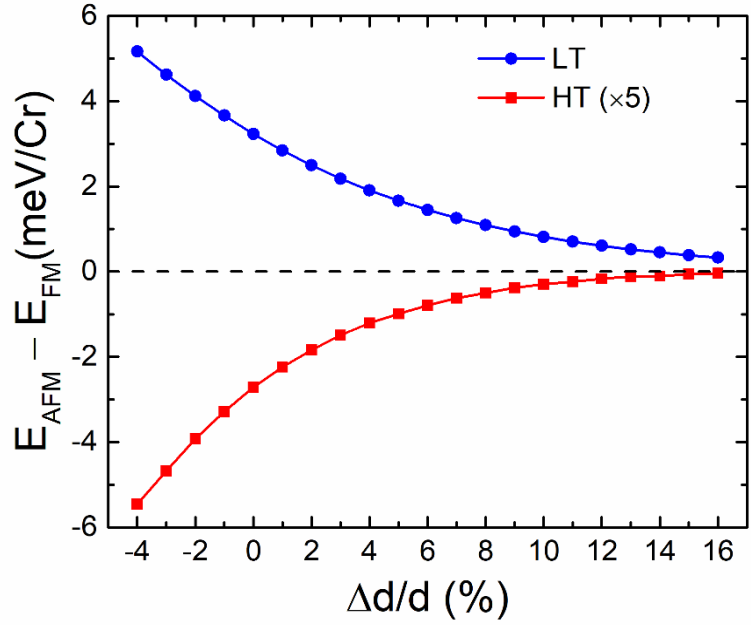

Figure S4 Energy differences between AFM and FM states ($E_{AFM}$-$E_{FM}$) under different layer distances for the HT and LT phases. The energy differences of the HT phase are enlarged 5 times. The magnetic configurations of both HT and LT phases CrI$_3$ bilayer will not be altered when the interlayer distance changes, which indicates the magnetic configurations are not sensitive to the uniaxial strain perpendicular to the *xy* plane. Spin-orbit coupling was considered here.

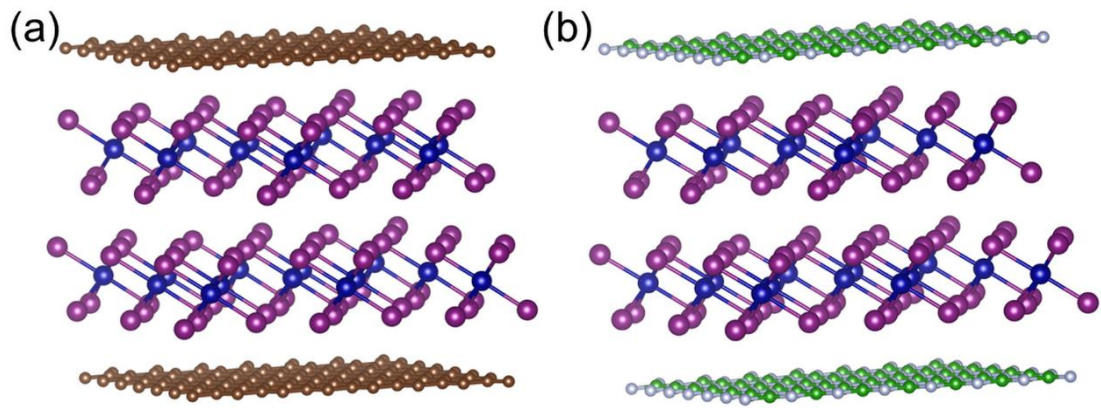

Figure S5 The structures of CrI$_3$ bilayer with graphene (a) and BN (b) substrates. We have added two graphene or BN monolayers over and below the CrI$_3$ bilayer respectively as substrates to simulate the environment influence, which are similar to the experimental structures (Science 360, 1214 (2018); Science 360, 1218 (2018); Nat. Nanotech. 13, 544 (2018); Nat. Mater. 17, 406 (2018)). Our calculations show that the substrates have no much influence to the magnetic configurations in both HT and LT phases. In addition, we also estimated the strain effect induced by the substrate, which also show no appreciable influence on our results.

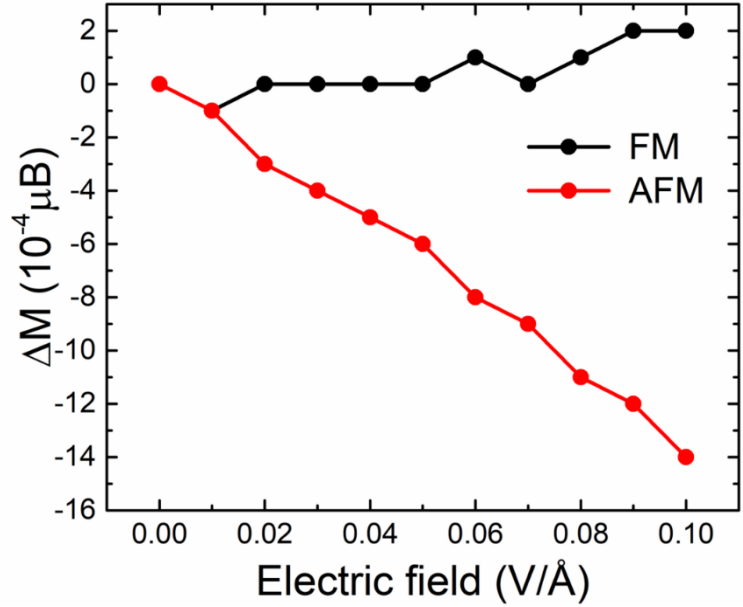

Fig. S6 Relative changes of total magnetization (4 Cr atoms included) of both FM and AFM configurations in the HT phase as a function of electric field. The total magnetization of its AFM ground state in the HT-phase increases slightly with respect to the electric field applied. This should be attributed to i) the charge transfer between two layers under electric field and ii) the converse magnetization direction of each layer. However, the total magnetization of the FM configuration in the HT phase (can be realized by doping or external magnetic field) is less influenced under electric field since the magnetization direction of each layer is exactly same. In Ref. 11, it was concluded that "in the AFM phase, the electric field E induces a constant magnetization that increases with E", and "in the FM phase, $M_0$ is nearly independent of E". All the statements are consistent with our theoretical calculations and thus verify the solidness of our conclusions. Please note our calculated variation of the total magnetization under electric field is smaller than that in Ref. 11, in which it should possible result from a combined effect of both electric field and charge doping. Unfortunately, the method combining the both effects is under developing, which is not directly relevant to the present work.

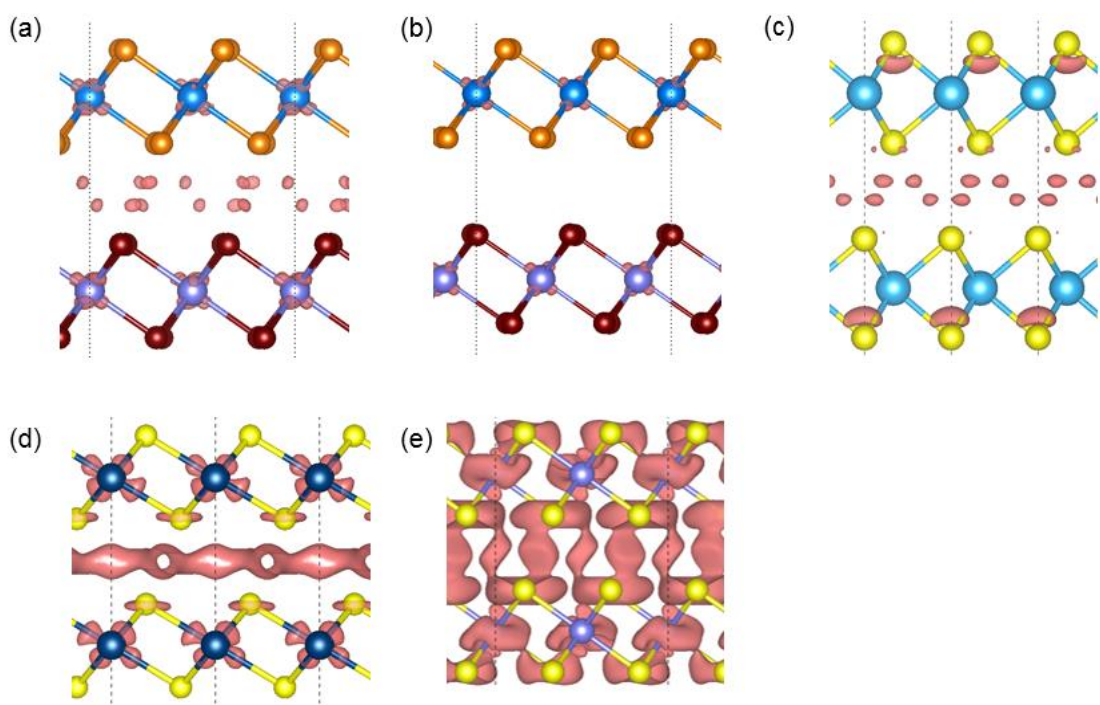

Figure S7 Differential charge densities in bilayer $CrI_3$-LT (a), $CrI_3$-HT (b), $MoS_2$ (c), $PtS_2$ (d) and $CrS_2$ (e), respectively. Here, red isosurface contours represent charge accumulation after layer stacking and the isosurface value is 0.002 e/ $Bohr^3$. Interlayer coupling in layered materials could redistribute charge density at the interlayer region. The amount of redistributed charge in $CrI_3$-LT is comparable with that previously found in $MoS_2$.

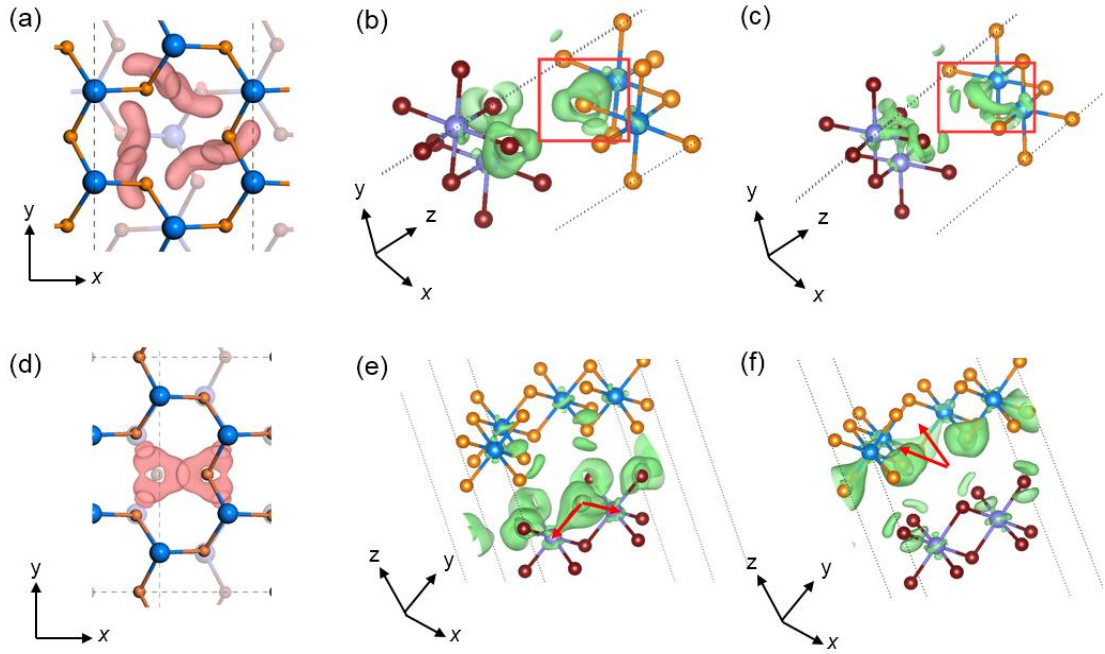

Figure S8 Spin-dependent differential charge densities in bilayer LT and HT CrI$_3$. To illustrate the charge transfer between two layers with different spin component, spin dependent differential charge densities are shown in Figure S8. Light-rose and green isosurface indicate charge accumulation and depletion after layer stacking, respectively. Figure S8(a) shows the charge accumulation after layer stacking in LT phase. The charge accumulation for different spin components shares an almost same pattern, thus is shown with only one figure, which is also quite similar with that of the total charge shown in the main text. Figure S8(b-c) show the charge reduction of spin up and down, respectively. For spin down (c) charge reduction mainly occurs around the $p_{x/y}$ orbitals of I atoms near the interlayer region, indicating a direct FM coupling between the spin-down components (green) of the two I atoms. In the HT case, charge sharing takes a different pattern as shown in the second panel in Figure S8. Charge transferred from I atoms forms a triangle-shaped charge accumulation in the interlayer region. Direct charge transfer from the $p_{x/y}$ orbital to the $p_z$ orbital is observable in the (e) and (f), indicating a direct $p_{x/y}$-$p_z$ interaction.

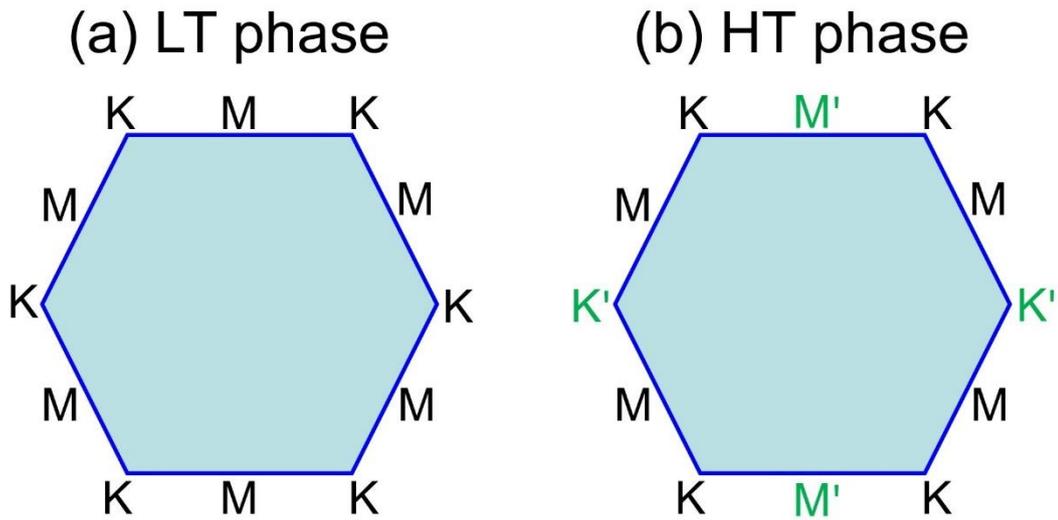

Figure S9 First Brillouin zone of LT (a) and HT (b) phases, respectively. For the LT bilayer, it exhibits the same stacking order with graphene bilayers where the hexagonal symmetry is maintained, hence the six M/K points keeps equivalent. However, the symmetry is reduced in the HT bilayer because of the sliding of the upper layer. As a result, the six M/K points in first Brillouin zone split into four M/K points and two M'/K' points.

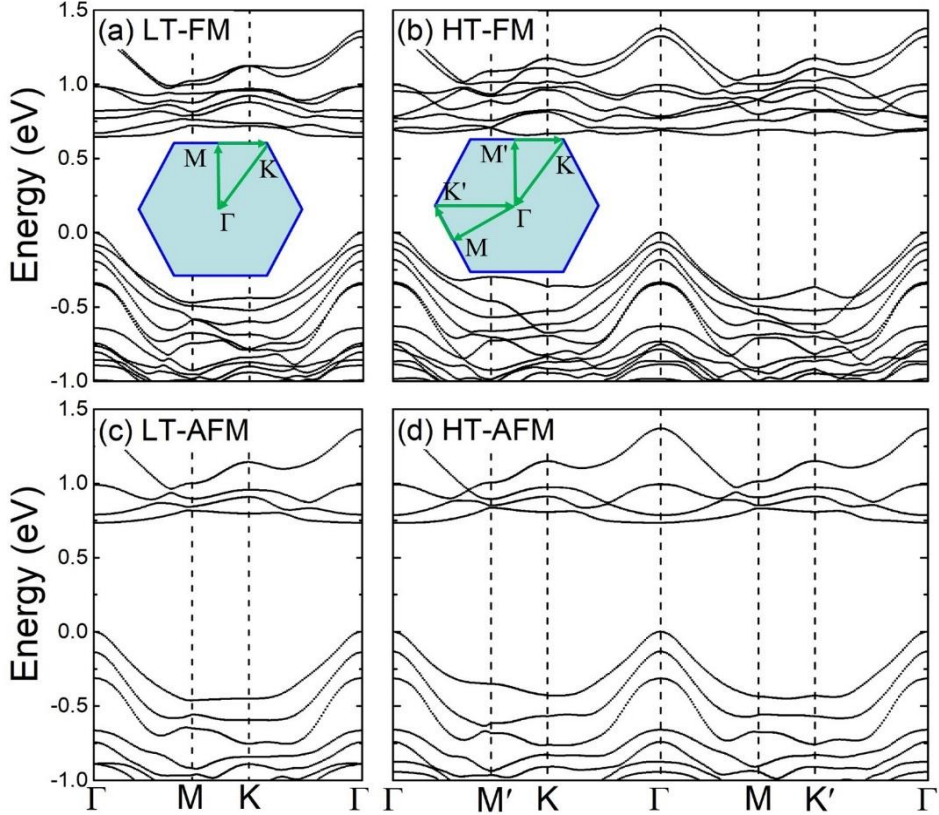

Figure S10 Band structures of bilayer CrI$_3$. (a) LT phase with interlayer FM state; (b) HT phase with interlayer FM state; (c) LT phase with interlayer AFM state; (d) HT phase with interlayer AFM state. Spin-orbit coupling has been considered in all calculations. The insets are first Brillouin zones of two phases, which show the M' and K' are nonequivalent with M and K in the HT phase (see detail in Figure S9). It should be noted that the band structures of the two AFM states are double degenerate while it is not the case for FM ones. The reduced symmetry in HT phase leads to appreciable variations of band structures. The highest valence band (HVB) at the nonequivalent M' point is upshifted compared with the M point in either HT-FM or HT-AFM, resulting in the flatter HVB of the HT bilayer than that of the LT bilayer. The PBE bandgaps of LT-FM, HT-FM, LT-AFM, and HT-AFM are 0.64 eV, 0.66 eV, 0.73 eV, and 0.73 eV, respectively. All of them, except HT-FM, are direct bandgap semiconductors. The indirect bandgap of HT-FM results from a downshift of the lowest conductance band along M'-K. We thus expect diverse electronic band structures for LT-FM and HT-AFM because of the nonequivalent symmetry, different degenerate situations between the FM and AFM states, and variant bandgaps (differs about 90 meV) between them, as shown in (a) and (d).

**Supplementary Table SI**

Relative total energy $\Delta E_0$ (with respect to the most stable configuration LT-FM), the binding energy, magnetic moment per Cr, lattice constants $a$, and the interlayer distance in a $CrI_3$ bilayer. The lattice constants show little difference with different phase and magnetic configuration. The binding energy of configuration LT-FM is -14.7 meV/Å$^2$, which is slightly smaller than those $PtS_2$ and $MoS_2$.

| Configuration | $\Delta E_0$ (meV/Cr) | $E_b$ (meV/Å$^2$) | Cr-Mag. ($\mu_B$) | I-Mag. ($\mu_B$) | $a$ (Å) | $d$ (Å) |
|---|---|---|---|---|---|---|
| LT-FM | 0 | -14.7 | 3.28 | -0.12 | 6.92 | 3.48 |
| LT-AFM | 2.9 | -14.4 | 3.27 | -0.12 | 6.92 | 3.49 |
| HT-FM | 3.4 | -14.4 | 3.28 | -0.12 | 6.92 | 3.44 |
| HT-AFM | 2.4 | -14.5 | 3.28 | -0.12 | 6.92 | 3.46 |
| PtS$_2$ | | -27.5 | | | | |
| MoS$_2$ | | -25.3 | | | | |

**Supplementary Table SII**

Energy differences between interlayer AFM and FM states ($E_{AFM}$-$E_{FM}$) with consideration of different exchange functionals (including HSE hybrid functional). The spin-orbit coupling has been considered in all calculations except the HSE one. All calculations give similar results which indicate our conclusion is quite robust and will not be changed under a different functional.

| $E_{AFM}$-$E_{FM}$ (meV/Cr) | PBE | PW91 | LDA | revPBE | PBEsol | HSE |
|---|---|---|---|---|---|---|
| HT | -0.544 | -0.815 | -0.823 | -0.554 | -0.698 | -2.295 |
| LT | 3.231 | 3.107 | 3.551 | 2.916 | 3.422 | 1.685 |

**Supplementary Table SIII**

Energy differences between interlayer AFM and FM states ($E_{AFM}$-$E_{FM}$) with consideration of different vdWs. All calculations give similar results which indicate our conclusion is quite robust and will not be changed under a different vdW.

| $E_{AFM}$-$E_{FM}$ (meV/Cr) | No vdW | IVDW | | | |
|---|---|---|---|---|---|
|  |  | DFT-D2 | DFT-D3 | TS |  |
| HT | -0.908 | -0.908 | -0.908 | -0.906 |  |
| LT | 2.825 | 2.825 | 2.825 | 2.826 |  |
| $E_{AFM}$-$E_{FM}$ (meV/Cr) | vdW-DF | | | | |
|  | revPBE | optPBE | optB88 | optB86b | vdW-DF2 |
| HT | -1.040 | -1.036 | -1.151 | -1.018 | -1.253 |
| LT | 2.636 | 2.787 | 2.877 | 2.936 | 3.031 |